A Science Whitepaper submitted in response to the 2010 Decadal Survey Call

# Mass Transport Processes and their Roles in the Formation, Structure, and Evolution of Stars and Stellar Systems

## 12 February 2009

Kenneth G. Carpenter (NASA-GSFC), Margarita Karovska (CfA), Carolus J. Schrijver (LMATC), Carol A. Grady (Eureka Scientific), Ronald J. Allen (STScI), Alexander Brown (UColo), Steven R. Cranmer (CfA), Andrea K. Dupree (CfA), Nancy R. Evans (CfA), Edward F. Guinan (Villanova U.), Graham Harper (UColo), Antoine Labeyrie (College de France), Jeffrey Linsky (UColo), Geraldine J. Peters (USC), Aki Roberge (NASA-GSFC), Steven H. Saar (CfA), George Sonneborn (NASA-GSFC), and Frederick M. Walter (SUNY)

For more Information, please contact:

Dr. Kenneth G. Carpenter
Code 667, NASA-GSFC
Greenbelt, MD 20771
Phone: 301-286-3453, Email: Kenneth.G.Carpenter@nasa.gov

*Introduction*

*Understanding the formation, structure, and evolution of stars and stellar systems remains one of the most basic pursuits of astronomical science, and is a prerequisite to obtaining an understanding of the Universe as a whole.* The evolution of structure and transport of matter within, from, and between stars are controlled by dynamic processes, such as variable magnetic fields, accretion, convection, shocks, pulsations, and winds. Future long-baseline (0.5-1.0 km) observatories (i.e., space-based interferometers and sparse aperture telescopes) will achieve resolutions of 0.1 milli-arcsec (mas), a gain in spatial resolution comparable to the leap from Galileo to HST. As a result, spectral imaging observations from such facilities will enable a quantum leap in our understanding of stars and stellar systems – but *only* if investments in technology development are made in the current decade to enable the launch of such missions in the following decade. In this whitepaper, we discuss the compelling new scientific opportunities for understanding the formation, structure, and evolution of stars and stellar systems that can be enabled by dramatic increases in UV-Optical angular resolution to the sub-mas level. An Ultraviolet-Optical Interferometer (UVOI) under consideration for the decade of the 2020s and beyond will provide direct spectral imaging of spatial structures and dynamical processes in the various stages of stellar evolution (e.g., Fig. 1) for a broad range of stellar types.

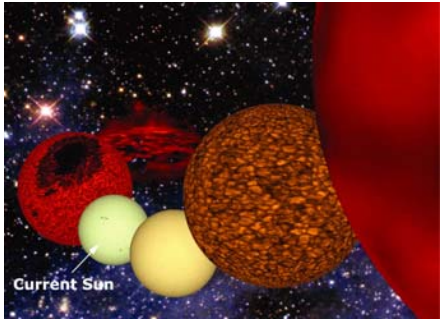
Fig. 1: Evolution of the Sun in time.

We discuss below the opportunities available for dramatically improved observation and understanding of: Young Stellar Systems; Hot Star rotation, disks, & winds; Stellar Pulsation across the HR-diagram and its impact on stellar structure and mass loss; convection in Cool, Evolved Giant and Supergiant Stars; Interacting Binaries; novae and Supernovae. Hours to weeks between successive images (see Fig. 2) will detect dramatic changes in many objects, e.g., mass transfer in binaries, pulsation-driven surface brightness variation and convective cell structure in giants and supergiants, jet formation and propagation and the changes in debris disks/shells in young planetary systems due to orbiting resonances and planets, non-radial pulsations in and winds from stars, and the structure, evolution, and interaction with the ISM of the core regions of nearby supernovae.

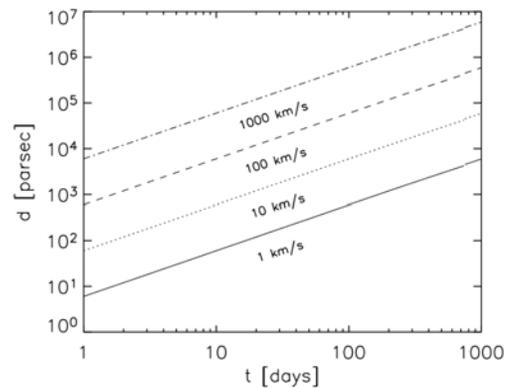
Fig. 2: Minimum time interval between successive images required to resolve the motion of a feature moving at different speeds, as a function of the object's distance.

*Dynamic Processes in Young Stellar Systems: Star Formation, Protoplanetary Disks and Jets*

Protoplanetary disks are where the materials that can ultimately produce life-bearing worlds are assembled. For our own Solar System, the first 50 Myr spans the formation and evolution of the proto-Solar nebula, the assembly of the meteorite parent bodies, the formation of the proto-Earth and proto-Mars, and the early phases of the Era of Heavy Bombardment. *If we are to understand not only the history of our Solar System, but also how planetary systems develop in general, we need to understand the disks, how long they last, how they interact with their central stars, and how they evolve.*

For the first few million years, both young solar type (T Tauri) and intermediate-mass (Herbig Ae) stars continue to accrete material from their disks. The inner boundaries of these disks are expected to be at the corotation radius from the star, typically 3-5 stellar radii (~0.05 AU for the T Tauri stars). The environment closer to the star is controlled by the strong stellar magnetic field, with accreting material channeled along field lines to the stellar photosphere. In the accretion shock plasma temperatures increase from several thousand to a few million degrees. Due to the high temperatures, UV emission from the chromosphere and the accretion spot(s) is detectable at high contrast against the lower-temperature stellar photosphere. While inner disk edges have been resolved by HST for dust disk cavities with radii in the 10-20 AU range[11], the inner edge of the gas disk has yet to be resolved for any young star with HST, but would be resolved with a UVOI for stars as distant as 160 pc. Fig. 3 shows a simulation of such an observation of the Lyα-fluoresced $H_2$ emission originating in the magnetosphere-disk interaction region of a T Tauri star at ~50 pc. Determining the size and geometry of the field-dominated region is of great importance for understanding stellar rotational braking, and accretion rates[3] as a function of global disk parameters. In addition to providing the size of the region, repeated observations may reveal rotation of resonances and indirectly point to the location of planets. Moreover, direct detection of planets associated with young, active stars may be possible via their UV auroral emissions or via transits and the impact of close-in exoplanets on the activity of their hosts stars may be evaluated.

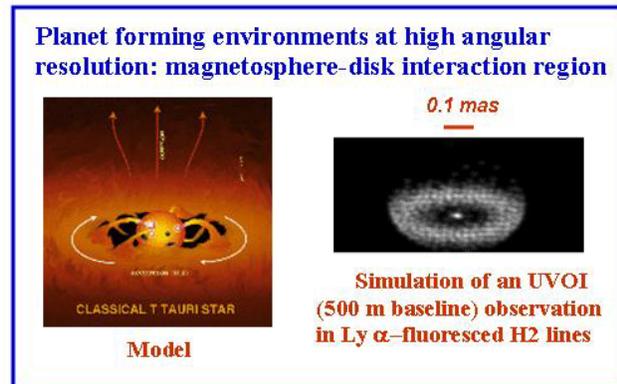

Fig. 3: A simulation of a sub-mas observation of the Lyα-fluoresced H2 emission originating in the magnetosphere-disk interaction region of a T Tauri star at ~50 pc.

Red-shifted absorption features in T Tauri star spectra[7] and lack of X-ray eclipses[12] has been interpreted as indicating that the accretion footprints on young stars are at high stellar latitudes. Sub-mas spatial resolution will allow us to directly image the accretion hot spot(s), and provide a map of the accretion flow from the co-rotation radius of the disk onto the accretion footprints, using emission lines spanning a wide ionization range. Such imagery will allow us to test how the accretion geometry changes with stellar mass, age, and disk properties.

Young stars frequently drive bipolar outflows that can be traced from a few arcseconds up to parsec scales[19]. For nearby stars, these outflows can be resolved from the central star in H Lyα and optical forbidden lines, and in some cases in Si III 1206.5 Å[11]. HST imagery suggests that by a few AU from the star, the bipolar outflows are orthogonal to the disk. This is unexpected for outflows that are solely following the stellar magnetic field geometry, since the stellar-rotation and global-magnetic axes are not perfectly aligned in the Sun, or for the few PMS stars studied in detail[6]. One possibility is that the outflow launch and collimation region extends over the inner few AU of the disk[27], but the detailed coupling between mass loss from the star and the larger-scale jet is currently poorly understood. Sub-mas observations in the UV would also allow us to resolve any uncollimated wind component. Such a wind component has been proposed as a means of transporting annealed silicates and processed organics from the inner parts of the protoplanetary disk into more distant icy planetesimals, thus accounting for the compositional diversity of comet nuclei[26].

*Dynamic Processes in Hot stars: Rotation, Disks, Winds, and Circumstellar Envelopes*
There are many competing processes on stars that produce structures on the surface or in the circumstellar environment. These processes include radiative winds, rapid rotation, pulsations, and magnetic fields, many of which may operate simultaneously within the stellar envelope.

*Understanding how massive stars rotate is important for the accurate modeling of stellar evolution and computing the final chemical yields of stars*[18]. Hot (O, B, Wolf-Rayet) stars tend to be the most rapidly rotating types of stars (excluding degenerate stars), and many are rotating so fast that their shapes are centrifugally distorted into oblate spheroids. Although rapid rotation in the very rare eclipsing binaries is measurable using light curves and radial velocity profiles, it is extremely difficult to pin down the detailed properties of single-star rapid rotation. A UVOI would enable direct measurement of the rotation rate and any differential rotation by imaging features moving across the star at different latitudes. Imaging the stellar oblateness will provide a better measure of the star's total angular momentum than feature-tracking alone could provide.

*Hot stars exhibit strong stellar winds that contribute significantly to the mass and energy balance of the interstellar medium.* Quantitative modeling of UV spectral features associated with stellar winds has evolved into a reasonably accurate means of deriving fundamental stellar parameters and distances[15]. The atmospheres and winds of hot stars are intrinsically variable, and it is now accepted that in many cases time-dependent phenomena (e.g., pulsations or magnetic field evolution) in the photosphere provide "shape and structure" to the wind[8]. The direct observational confirmation of a causal connection between specific stellar variations and specific wind variations, though, has proved elusive. For many O and B stars, it is not clear whether large-scale wind inhomogeneities are rotationally modulated (i.e., due to spots) or if pulsations are responsible, or if the variability occurs spontaneously in the wind. Sub-mas observations would shed light on the origins of wind variability. Simply seeing correlations between individual spots (no matter their physical origin) and modulations in the wind would be key to understanding how hot stars affect their local environments. One paradigm to be tested is the idea that discrete absorption components (DACs) are caused by corotating interaction regions (CIRs) in the winds[5]. While continuum-bandpass filters can be used efficiently to search for thermal and diffusive inhomogeneities on a hot star's disk, most other processes are best studied by imaging in UV spectral lines. From the ground one can do some imaging in Hα, but it is so optically thick that structures are hard to see. In the UV, however, the CIV doublet can be employed to study inner winds and co-orbiting structures of hot stars, while the MgII doublet can be used to trace the discrete ejections of mass and the extent of disks out to several stellar radii.

Classical Be stars are ostensibly-single, rapidly-rotating, probably post-Zero Age Main Sequence (ZAMS) stars which eject mass that episodically collapses onto an equatorial disk. The observed properties of Be stars and their circumstellar gas are consistent with the coexistence of a dense equatorial disk and a variable stellar wind[22]. One of the longest-standing puzzles in hot-star astrophysics is the physical origin of this disk, both from the standpoint of mass supply (the winds may be too tenuous) and angular momentum and energy supply (the disk particles are in Keplerian orbits but the stellar surfaces are not). *Direct UV/Optical imaging of Be stars will provide answers regarding the physical distribution of matter, structures within the disks and winds (spiral density waves or clumpy structures), wind/disk interaction regions, and ionization structure, and the variation of these parameters with time.* Sub-mas resolution would allow characterization of the mean disk properties (e.g., inclination, radial density structure, and thickness) as a function of spectral type and stellar rotation rate, which would provide stringent empirical constraints on the wide variety of proposed, but as yet unverified, theories.

Wolf-Rayet (WR) stars are believed to be the central, heavy cores of evolved O-type stars that have lost most of their hydrogen-rich outer layers as a stellar wind. *WR stars have observed mass loss rates at least an order of magnitude higher than other O stars (i.e., of order $10^{-4}$ $M_{sun}$/yr), and the origin of these extremely dense and optically thick outflows is still not well understood.* The only way that line-driven wind theory can account for such large mass loss rates is if the opacity in the lines is utilized many times (i.e., if photons multiply scatter through the optically thick outer atmosphere before they give up all of their radiative momentum to the gas[9]. However, other ideas exist, including fast magnetic rotation[13] and "strange-mode" pulsations in the chemically enriched interiors[10]. Direct imaging of winds at all levels, including dense clumps would permit direct evaluation of these models, and an assessment of the role of turbulence[17] in these massive winds. Turbulence greatly complicates the study of line-driven mass loss[2] and makes it much more difficult to understand how WR winds are accelerated by radiation. Sub-mas UV observations are needed in order to identify the origin and nature of these clumps and clarify the physical processes responsible for these winds.

*Pulsation Processes and their Impact on Stellar Structure and Mass Loss*

Pulsations are found in many different types of stars, ranging from very hot main-sequence stars to dying cool giants and supergiants, and stellar relics. *In many cases stellar pulsations, radial or non-radial, significantly affect the extent, composition, and structure of stellar atmospheres.* The signatures of pulsation are very prominent in the UV (e.g. Mg h&k lines) and a UVOI will enable direct imaging of pulsation effects including surface structures and shock fronts as they propagate through the dynamical atmospheres. Images of the effects of the pulsation will provide key inputs to hydrodynamical models for a range of diverse pulsators, such as Miras and Cepheids, cool supergiants, and hot B-stars. Direct observation of the shock-propagation in extended stellar atmospheres and winds will characterize the time evolution and spatial symmetries of shocks and constrain and improve theoretical shock models in stars with a wide range of masses. These observations will answer a large number of crucial questions about stellar interiors, core convection, chemical mixing, and magnetic fields.

Nonradial pulsations (NRP's) produce evenly-spaced temperature modulations that can be imaged as bright and dark zones on the star. Surface thermal modulations may amplify wind flows into clumps. The ultimate tests of both interior pulsation theory and line profile models will be the counting of the hot/cool zone pairs on the star and the determination of whether they only are concentrated on a star's equator. Theories of NRPs, e.g., in very rapidly rotating stars, are still evolving, and the imaging of how rotation affects the latitudinal profile of pulsation amplitudes would verify or falsify certain modeling assumptions and directly diagnose the angular momentum profiles of these stars[28]. *For example, the direct imaging of a cause-and-effect relationship between stellar and circumstellar features could provide the long-sought explanation for the Be phenomenon.*

β Cephei stars are pulsating variables which generally possess radial or low-degree nonradial modes and therefore show a coherent variation over the surface at any given phase. The pulsations in some of these stars (e.g., BW Vul) are the largest in terms of velocity amplitude of any known variable class. In some pulsators the resulting shocks form transient shells which could be imaged in lines of CIV as they emerge from the surface, coast to a standstill, and subsequently return to the star in near free fall, enabling a determination of whether mass loss can be driven effectively by this piston-type interaction. Direct observation of the spatial structure of these pulsations will identify the pulsation modes and provide valuable information about the internal properties of these stars which are progenitors of supernovae.

*Convection in Cool Evolved Giant and Supergiant Stars*

Stars that are at least 1.5x heavier than the Sun are not magnetically active during their mature life on the main sequence because they lack envelope convection. Consequently, they begin their transformation to red giant stars with essentially the same rotational energy they had after their initial formative epochs. As they expand, a dynamo is activated once the star cools enough to develop envelope convection. That may lead to significant, sudden magnetic braking, which possibly results in a substantial difference between the rotation rates of the deep interior and the magnetically-active convective envelope[25]. Observations indicate that such a difference may last for up to some tens of millions of years. Detailed understanding of the onset of dynamos in evolving stars with such shear layers between envelope and interior, and of the possible consequences for the internal dynamics, will greatly benefit from imaging and disk-resolved seismic observations of stars in such evolutionary phases.

Continuing their evolution as red giants, the stars reach a point where the coronal activity disappears again, to be replaced by substantial mass loss at much lower temperatures. In a HR-diagram this behavior occurs on either side of a dividing line. Even though there is an absence of magnetically heated transition-region and coronal plasma in the late-K and M-type giant stars, their winds are thought to be driven by magneto-hydrodynamic waves. It has been proposed[24] that the coronal dividing line is a consequence of a dynamo transition: large-scale structures with closed field lines and coronal heating, and small-scale structures with open field lines and increased mass loss. The hybrid stars that display both phenomena are the key to understanding the dividing line and the associated change in the dynamo mode from global to local. Sub-mas imaging of the transition-region and chromospheric emissions in the UV will reveal the magnetic field topology on stars on both sides of the dividing line, and on the hybrid-atmosphere stars.

As stars expand to supergiant stages, the scale of the surface convection changes to the point that we expect only a few convective 'granules' to cover the entire star. Fig. 4 shows a model and simulated sub-mas observation of this convection. Does this really happen? Some doubt it because the spectral lines of these stars show little sign of such large-scale turbulence. And if it does, then a turbulent local dynamo may again create magnetic fields on a near-global scale. A UVOI can image both the large-scale convection (and its evolution) and possible chromospheric patterns driven by this process.

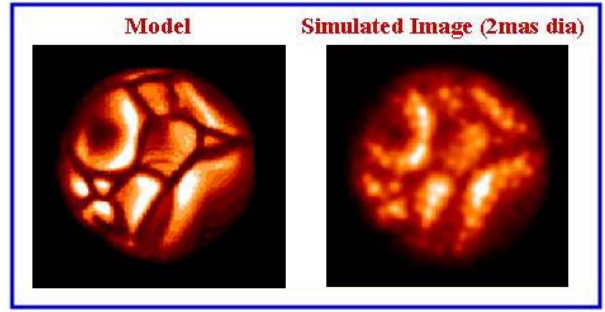

Fig. 4: Model (B. Freytag) and simulated observation (500m baseline) of the convection on a supergiant like α Ori at 2 kpc. These "granules" transport the energy from the interior to the surface, evolving on a timescale of years, with ~dozen granules filling the entire surface.

*Interacting Binary Systems: Understanding Accretion Processes*

*Almost all high-energy sources in the Universe are powered by potential energy released via accretion.* Understanding accretion driven flows in binaries will directly affect our understanding of similar flows around YSOs, including the formation of planets in the circumstellar disk, as well as the much larger scale accretion flows in active galactic nuclei (AGNs). Compact, mass transferring binaries provide us with laboratories for testing energetic processes such as magnetically driven accretion and accretion geometries, and various evolutionary scenarios.

In close binary stars the flow of material from one component into the potential well of the other determines the future evolutionary histories of each component and the system itself, and particularly the production of degenerate companions and supernovae. Our cosmological standard candles, the Type Ia supernovae, for example, may be a consequence of accretion onto a white dwarf in a close binary. Currently, most of our accretion paradigms are based on time-resolved spectroscopic observations. For example, in Cataclysmic Variables (CVs) the picture of accretion onto compact objects via an extended accretion disc is solidly based on spectral and timing information. However, several objects challenge our standard picture and there are significant gaps in our understanding of their formation and evolution.

Large uncertainties exist in our quantitative understanding of accreting processes in many interacting systems. The interaction between the components in close binaries is believed to occur via Roche lobe overflow and/or wind accretion. 3-D hydrodynamic simulations show that the accretion processes in tidally interacting systems are very complex. Wind accretion is even more complicated. The amount of the accreted material depends on the characteristics of both components including stellar activity and wind properties (e.g. density and velocity), the binary parameters (e.g. orbital period and separation), and the dynamics of the flow. In the case of Roche lobe overflow, the accretion may form an extended accretion disk whose turbulent magnetic dynamo drives the flow through it and may launch collimated outflows and jets. Stellar activity of the rapidly rotating donors and their impact on the binary remains poorly constrained despite being crucial in regulating the mass transfer rate and setting the long-term evolution.

One key to further advances in accretion studies is resolving a wide range of interacting binaries and studying their components and mass flows. Sub-mas resolution in the UV will lead to unprecedented opportunities for detailed studies of accretion phenomena in many interacting systems, including symbiotics[14], Algol-type binaries (Fig 5), and CVs[20]. A UVOI will be able to resolve the components of numerous interacting systems and will therefore provide a unique laboratory for studying accretion processes and jet-forming regions. The binary components can be studied individually at many wavelengths including Lyα, NV, CIV, and MgII h&k lines, and the geometry of accretion, including high temperature regions, hot accretion spots[21], bipolar flows and jets, and can be imaged directly, giving us the first direct constraints on the accretion geometries. This in turn will allow us to benchmark crucial accretion paradigms that affect any stellar population and even the structural evolution of galaxies whose central black-holes are steadily accreting, shaping their long term evolution.

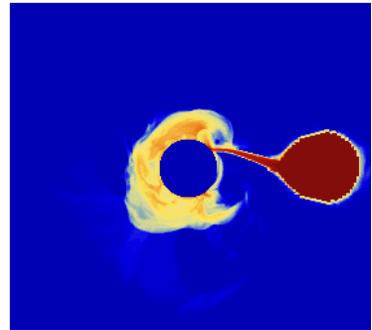

Fig. 5: Hydrodynamic simulations[23] of the mass transfer in the Algol prototype β Per (2 mas separation), showing H-alpha emissivity. The gas stream impacts onto the surface of the primary forming a local hotspot as well as an extended flow around the accretor.

*Supernovae and Novae*

With the exception of the relatively nearby SN1987A (in the LMC), which could be well-studied by HST, it has not been possible to obtain much information about the close-in spatial structure of supernovae (typical sizes remain below about 1 mas, which is not reached by current ground-based optical telescopes). Radio VLBI observations have resolved a few supernovae, but are more a probe of the interaction of the SN shock front with the circumstellar material than of the supernova[1]. Direct imaging at the sub-mas level would resolve early stages of expansion of supernovae at a distance of few Mpc, and of galactic novae. These images would provide essential information on the nature of the explosion, especially in regard to its symmetry or asymmetry, and of the early evolution of its structure with time.

*Conclusion*

We have summarized some of the compelling new scientific opportunities for understanding stars and stellar systems that can be enabled by sub-mas angular resolution, UV/Optical spectral imaging observations, which can reveal the details of the many dynamic processes (e.g., variable magnetic fields, accretion, convection, shocks, pulsations, winds, and jets) that affect their formation, structure, and evolution. These observations can only be provided by long-baseline interferometers or sparse aperture telescopes in space, since the aperture diameters required are in excess of 500 m – a regime in which monolithic or segmented designs are not and will not be feasible - and since they require observations at wavelengths (UV) not accessible from the ground. Two mission concepts which could provide these invaluable observations are NASA's Stellar Imager[4] (SI; http://hires.gsfc.nasa.gov/si/) interferometer and ESA's Luciola[16] sparse aperture hypertelescope, which each could resolve hundreds of stars and stellar systems. These observatories will also open an immense new discovery space for astrophysical research in general and, in particular, for Active Galactic Nuclei (Kraemer et al. Science Whitepaper). The technology developments needed for these missions are challenging, but eminently feasible (Carpenter et al. Technology Whitepaper) with a reasonable investment over the next decade to enable flight in the 2025+ timeframe. That investment would enable tremendous gains in our understanding of the individual stars and stellar systems that are the building blocks of our Universe and which serve as the hosts for life throughout the Cosmos.